\documentclass[aps,prl,twocolumn,groupedaddress,showpacs]{revtex4}

\usepackage{amssymb,amsfonts,amsmath}
\usepackage{bm}
\usepackage{graphicx}
\usepackage{subfigure}

\begin{document}
\title{Pitchfork bifurcations in blood-cell shaped dipolar Bose-Einstein condensates}

\author{Stefan Rau}
\author{J\"{o}rg Main}
\author{Patrick K\"{o}berle}
\author{G\"{u}nter Wunner}

\affiliation{Institut f\"{u}r Theoretische Physik 1, Universit\"{a}t Stuttgart, 70550 Stuttgart, Germany}

\date{\today}

\begin{abstract}
We demonstrate that the  method of coupled Gaussian wave packets is a full-fledged
alternative to direct numerical solutions of the Gross-Pitaevskii equation 
of condensates with electromagnetically induced attractive $1/r$ interaction,
or with dipole-dipole interaction. Moreover, Gaussian wave packets are superior in that they are capable of 
producing both stable and unstable stationary solutions, and thus of giving 
access to yet unexplored regions of the space of solutions of the 
Gross-Pitaevskii equation. We apply the method to clarify  the 
theoretical nature 
of the collapse mechanism of blood-cell shaped dipolar condensates:  On the route to collapse the condensate passes through a pitchfork bifurcation, where 
the ground state itself turns unstable,
before it finally vanishes in a tangent bifurcation.

\end{abstract}


\maketitle

Bose-Einstein condensates (BECs) with dipole-dipole interaction have 
become an active and exciting field of research because they offer the 
possibility of tuning the relative strengths of the short-range
isotropic contact interaction and the anisotropic long-range dipole interaction
by manipulating the $s$-wave scattering length via
Feshbach resonances, and thus of studying a
wealth of new phenomena that occur as one crosses the whole range from
dominance of the contact interaction to that of the dipole
interaction. The experimental realization of a BEC of chromium atoms
 \cite{EXGriesmaierPRL94,EXStuhlerPRL95,Bea08},
which possess a strong magnetic  dipole moment, has given additional
impetus to the field (for a comprehensive list of references see
the recent review by \textcite{Lahaye2009}). In the
dilute limit, the theoretical description of these condensates can be 
done in the framework of the Gross-Pitaevskii equation (GPE). 
This nonlinear Schr\"odinger equation has been solved in the literature 
so far by simple variational ansatzes, where the mean-field energy is 
minimized, e.g., with the conjugate gradient method or by imaginary time 
evolution. 
In this Letter we will show that the method of coupled Gaussian wave packets 
is an adequate alternative to solving the GPE of BECs with long-range
interactions. Moreover, we will show that the method is superior
in that it also yields unstable stationary solutions, and thus opens access 
to  regions  of the space of solutions of the GPE unexplored heretofore. 
As an application of   the method we will 
analyze in detail the theoretical nature of the collapse mechanism
of dipolar BECs.

The GPE for ultracold gases with long-range interactions, described
by the interatomic potential  $V_{\rm lr}(\bm{r})$, has the form
\begin{eqnarray}
 {i}\frac{\mathrm{d}}{\mathrm{d}t} \psi(\bm r,t) = \big[ -\Delta + \gamma_x^2 x^2 + \gamma_y^2 y^2 + \gamma_z^2 z^2
    \nonumber \\ + 8\pi N a \left | \psi(\bm{r},t)\right |^2 
   +V_{\rm lr}\left(\bm{r}\right)\big]  \psi(\bm r,t) \, ,
\label{eq:extended_GP}
\end{eqnarray}
where for dipolar interaction we have 
\begin{equation}
   V_{\rm lr}(\bm{r}) =  N \int \mathrm{d}^{3}\bm{r'}\frac {1-3\cos^2 \vartheta'} {|\bm{r}-\bm{r'}|^3}|
 \psi(\bm{r'},t)|^{2}.
\label{dipolarinteraction}
\end{equation}
with $\vartheta'$ the angle between $\bm{r}-\bm{r'}$ and the axis of an 
external magnetic field.
For completeness we will also consider the case of an isotropic 
``gravity-like'' attractive $1/r$ long-range interaction, 
\begin{equation}
   V_{\rm lr}(\bm{r}) =  - 2  N \int \mathrm{d}^3 \bm{r}' \frac{\left | \psi(\bm{r}',t)\right |^2}
   {\left | \bm{r} - \bm{r}' \right |} \,.
\label{monopolarinteraction}
\end{equation}
According to \textcite{ODellPRL84} this interaction could be electromagnetically 
induced by exposing the condensate atoms to an appropriately arranged set of 
triads of laser beams.  The appealing feature of such 
``monopolar'' condensates is that they can be self-trapping, i.e.\ exist 
without an external trapping potential.
The equations above have been brought into dimensionless form by introducing
natural units, which for monopolar interaction  ($V_{\mathrm{mono}}= -u/r$) are \cite{PapadopoulosrevA76,holgerPRA77,holgerPRA78} the  ``Bohr radius'' $a_{u}=\hbar^2/(m u)$ for lengths, the ``Rydberg energy'' $E_u=\hbar^2/(2 m a_u^2)$ for
energies and $\omega_u= E_u/\hbar$ for frequencies.
Natural units for dipolar atoms with magnetic moment $\mu$ are \cite{gelbPatrick} the dipole length  
$a_{\rm d} = \mu_0 \mu m/(2\pi \hbar^2)$, the dipole energy  $ E_\mathrm{d} = \hbar^2 / (2 m a_{\rm d}^2)   $
and the dipole frequency    $\omega_\mathrm{d} = E_\mathrm{d} / \hbar $.
The  quantities $\gamma_{x,y,z}$ in (\ref{eq:extended_GP}) denote  the trapping frequencies in the 
three spatial directions measured in the respective frequency units, $N$ is the number of bosons, and
$a$ the scattering length in units of $a_u$ and $a_\mathrm{d}$, respectively.

The most obvious way of solving the Gross-Pitaevskii equation (\ref{eq:extended_GP}) is its direct 
numerical integration on multi-dimensional grids using, e.g.,  fast-Fourier techniques. The {\em stationary ground state} can be obtained by imaginary time evolution. These calculations,  however, may turn out laborious, and physical insight can often
be gained using approximate, in particular, variational solutions. A common approach employed for 
determining the dynamics and stability of condensates both with contact interaction only  \cite{PerezPRL,PerezPRA} and with
additional long-range interaction  \cite{holgerPRA78}  is to assume a simple Gaussian form
 of the wave function, with time-dependent width parameters and phase, and to investigate the dynamics of these quantities. 
For dipolar condensates improvements on the simple Gaussian form were made by multiplying it by second-order Hermite polynomials  \cite{Ronen}. 

As an alternative to numerical quantum simulations on multidimensional grids we will 
extend the variational calculations in such 
a way that numerically converged results are obtained with 
significantly reduced computational effort compared to the exact quantum simulations but with 
similar accuracy. The method is that of {\em coupled Gaussian wave packets}. It was originally proposed by Heller  \cite{hellerJCP65,hellerJCP75} to describe quantum dynamics in the 
semiclassical regime, and was successfully applied to the dynamics of molecules and atoms in external fields
 \cite{FabcicIII}. The idea is to choose trial wave functions which are {\em superpositions} of $N$ different Gaussians
centered at the origin
\begin{equation}
 \psi (\bm{r},t) = 
\sum_{k = 1}^{N} e^{i\left(a_x^{k}x^{2}+a_y^{k}y^{2}+a_z^{k}z^{2}+\gamma^{k} \right)}
\equiv \sum_{k = 1}^{N}g^{k}(\bm{a}^k,\gamma^k;\bm{r})
\label{eq:psi_ansatz}
\end{equation}
where both the  width parameters $\bm{a}^k$ and the scalars $\gamma^k$ are complex quantities, with the
latter determining the weight and the phase of the individual Gaussian.
Inserting the ansatz (\ref{eq:psi_ansatz}) into the time-dependent
Gross-Pitaevskii equation and applying the time-dependent variational principle where 
 $||i \Phi(t) - H \psi(t)||^2  $ is minimized by varying $\Phi$, and afterwards $\Phi$ is set equal to  $\Phi = \dot \psi$,
 yields a set of ordinary differential equations for the
width parameters $\bm{a}^k$ and the scalars $\gamma_k$  
(cf. \cite{FabcicIII})
\begin{subequations}\label{equationsofmotion}
 	\begin{align}
 	 \dot a_{\beta}^k &= - 4 (a_{\beta}^k)^2 - \frac{1}{2} V_{2,\beta}^k; \qquad\qquad \beta = x,y,z; \\
	 \dot \gamma^k &= 2 i (a_x^k + a_y^k + a_z^k) - v_0^k.
	\end{align}
 \end{subequations}
The quantities $(v_0^k,V_2^k)$ with $k = 1,\hdots,N$ constitute the solution vector to the set of linear equations
\begin{eqnarray}
\sum_{k=1}^{N} \left\langle g^l |x_\alpha^m x_\alpha^n v_0^k |g^k \right\rangle + \frac{1}{2} 
\sum_{k=1}^{N} \left\langle g^l |x_\alpha^m x_\alpha^n \textbf{x}V_2^k\bm{x}|g^k \right\rangle  \nonumber \\
= \sum_{k=1}^{N} \left\langle g^l |x_\alpha^m x_\alpha^n V(\bm{x})|g^k \right\rangle
\end{eqnarray}
with $l=1,\hdots,N $; $m + n = 0,2$; and $x_1 =x$, $x_2 = y$, $x_3 = z$. 
Here, $V(\bm{x}) =  V_\mathrm{c} + V_{\rm lr} + V_\mathrm{t}$  denotes the sum of the contact, the long-range 
and of external trap potentials. The important and appealing point of this procedure is that 
all necessary integrals
with the trial wave functions $g^l, g^k$ from (\ref{eq:psi_ansatz}) can be calculated analytically.\\
Stationary variational solutions to the extended Gross-Pitaevskii equation (\ref{eq:extended_GP}) are found by searching for the fixed points of (\ref{equationsofmotion}), i.e. solving 
$\bm{\dot a}^k = 0; 
\dot \gamma^k = 0 $
for each $k = 1,...,N$ via a $4 N$ dimensional highly nonlinear root search. 
The resulting stationary width and weight/phase parameters can then be used to calculate the mean field energy 
$E_{\mathrm{mf}} = \left\langle  \Psi|-\Delta + V_\mathrm{t} + \frac{1}{2} (V_{\rm c} + V_{\mathrm{lr}}) |\Psi \right\rangle $ and the chemical potential 
$\mu = \left\langle  \Psi|-\Delta + V_\mathrm{t} + V_{\rm c} + V_{\mathrm{lr}} |\Psi \right\rangle $. It is important to note
that in contrast to numerical calculations with imaginary time evolution, which only work for
stable solutions, this procedure will produce both stable and
unstable solutions, and thus uncover yet unexplored parts 
of the space of solutions of the Gross-Pitaevskii equation.\\

To analyze the stability of the stationary solutions 
 the dynamical equations (\ref{equationsofmotion})
 are split into real ($R$) and imaginary ($I$) parts and  linearized around the fixed points. The eigenvalues of the Jacobian matrix \textbf{J} at the fixed point
\begin{equation} \label{Jacobian}
  \bm{J} = \frac{\partial 
 \big( \dot a_{\alpha}^{k,R}, \dot a_{\alpha}^{k,I},   \dot \gamma^{k,R}, \dot \gamma^{k,I}\big)
}{ \partial
\big(a_{\beta}^{l,R}, a_{\beta}^{l,I},  \gamma^{l,R},  \gamma^{l,I}\big)
}\,,
\end{equation}
with $\alpha, \beta = x,y,z;~k,l = 1,...,N$, determine the stability properties of the solution. If all eigenvalues $\lambda_j$ of the system are purely imaginary, the motion is confined to the vicinity of the fixed point and quasi-periodic. 
If one real part or several real parts of the eigenvalues are non-zero, small variations from the fixed point lead to exponential growth of the perturbation.

\begin{figure}[ht!]
\includegraphics[angle=0, width=0.9\columnwidth]{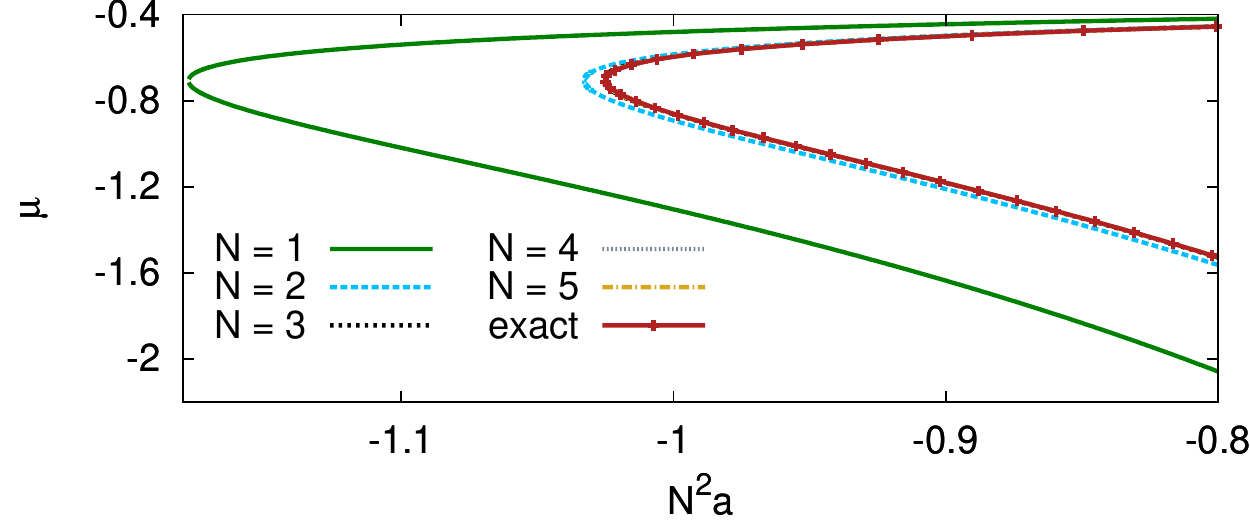}
\caption{\label{mumono} Chemical potential $\mu$ for self-trapped condensates with attractive
$1/r$ interaction as a function of the scaled scattering length $N^2 a$ obtained by using up to 
5 Gaussian wave packets in comparison with the result of the exact numerical solution of the stationary Gross-Pitaevskii equation.
Note that all forms yield a  tangential bifurcation diagram,  with a stable (upper) and an unstable (lower) branch.
The inclusion of three Gaussians already well reproduces the exact numerical result.}
\end{figure}

As a first application we demonstrate the efficiency of the coupled
Gaussian wave packet method for condensates with attractive $1/r$ long-range 
interaction, for the case of self-trapping ($\gamma_{x,y,z}=0$). 
Figure~\ref{mumono} shows the results for the chemical potential as 
a function of the scaled scattering length $N^2a$ for superpositions of 
1 to 5 Gaussians in comparison with the results of the exact numerical
solution. It is evident that all forms reproduce the bifurcation behavior 
discussed in  \cite{PapadopoulosrevA76,holgerPRA77,holgerPRA78}: at a critical point two solutions 
of the Gross-Pitaevskii equation are born in a tangent bifurcation, one stable (upper branch) and one
unstable (lower branch). The numerically accurate bifurcation point lies at $a_{\mathrm{cr}} \approx -1.025147 $.
 It can also be seen that, while the variational calculation with one Gaussian, with $a_{\mathrm{cr}}^{\rm N=1} = -1.178$, 
still lies far off the correct result, the inclusion of only one more Gaussian brings the chemical potential
curve already close to the numerical result, and practically no improvement is visible in Fig.~\ref{mumono} when
3 or more Gaussians are included. Using 5 coupled Gaussians the exact bifurcation point is reproduced with an accuracy of $10^{-6}$. Similar  results are obtained in the presence of a trapping potential.

We now turn to dipolar condensates. 
Previous studies \cite{Ronen, duttadelle} have shown that in certain regions of the parameter space dipolar
condensates assume a  non-Gaussian biconcave ``blood-cell-like'' shape. To demonstrate the power of the coupled Gaussian 
wave packet method, we choose a set of such parameters. We consider an axisymmetric trap with
 (particle number scaled) trap frequencies $N^2\gamma_z = 25200$ along the polarization direction of the dipoles and 
$N^2\gamma_{\rho} = 3600$ in the plane perpendicular to it (corresponding to an aspect ratio of 
$\lambda = {\gamma_z}/{\gamma_{\rho}}$ = 7). 
\begin{figure}[ht!]
\includegraphics[width=0.9\columnwidth]{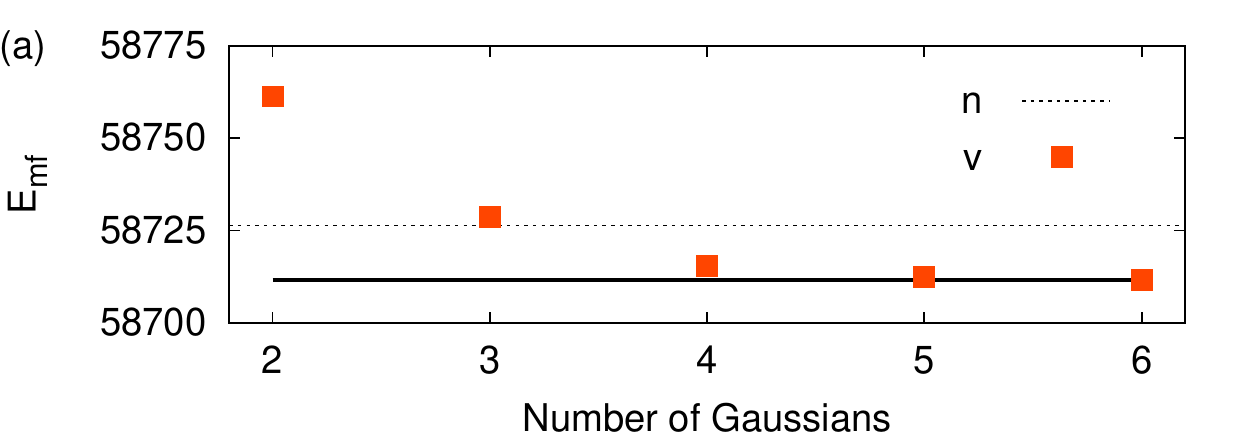}\\
\includegraphics[width=0.9\columnwidth]{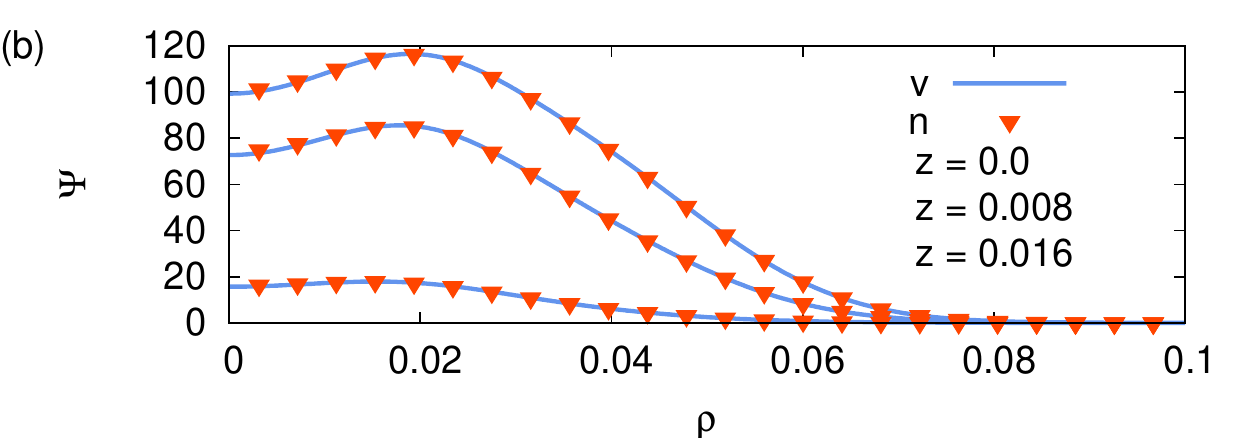}
\caption{\label{compare} (a) Convergence of the mean-field energy with increasing number of 
coupled Gaussian wave packets (squares) and comparison with the value obtained by a lattice calculation with grid size $128 \times 512$ (dashed line), which lies energetically higher than the exact converged variational solution (solid line). 
(b) Comparison  of the variational wave function for 6 coupled Gaussians 
(solid curves) with values of the numerical one  (triangles)
at different $z$ coordinates. Both solutions show a biconcave shaped condensate. The figures are for (particle number scaled) trap frequencies 
$N^2\gamma_z = 25200$ and  $N^2\gamma_{\rho} = 3600$, and scattering length $a = 0$.
}
\end{figure}
For this set of parameters we show in Fig.~\ref{compare}~(a) the convergence 
behavior of the mean field energy. 
We compare the variational solution as the number of Gaussian wave packets 
is increased from 2 to 6 with the mean field energy value of a numerical 
lattice calculation (imaginary time evolution combined with FFT) with a grid 
size of $128 \times 512$, at scattering length $a = 0$ as an example.
The mean field energy for one Gaussian is 
$E_\mathrm{mf} = 60361 \, E_{\rm d}$
and lies far outside the vertical energy scale. 
Evidently the numerical value is more than excellently reproduced by 5 and 6 
coupled Gaussians. 
The behavior for other scattering lengths is similar. Also the wave function nicely converges, and moreover, as can be seen in Fig.~\ref{compare}~(b), reproduces the biconcave shape of the condensate as does the numerical solution. Thus the method of coupled Gaussians is a viable and full-fledged alternative to direct numerical solutions of the Gross-Pitaevskii 
equation for dipolar condensates.
\begin{figure}[ht!]
 {\includegraphics[width=0.9\columnwidth]{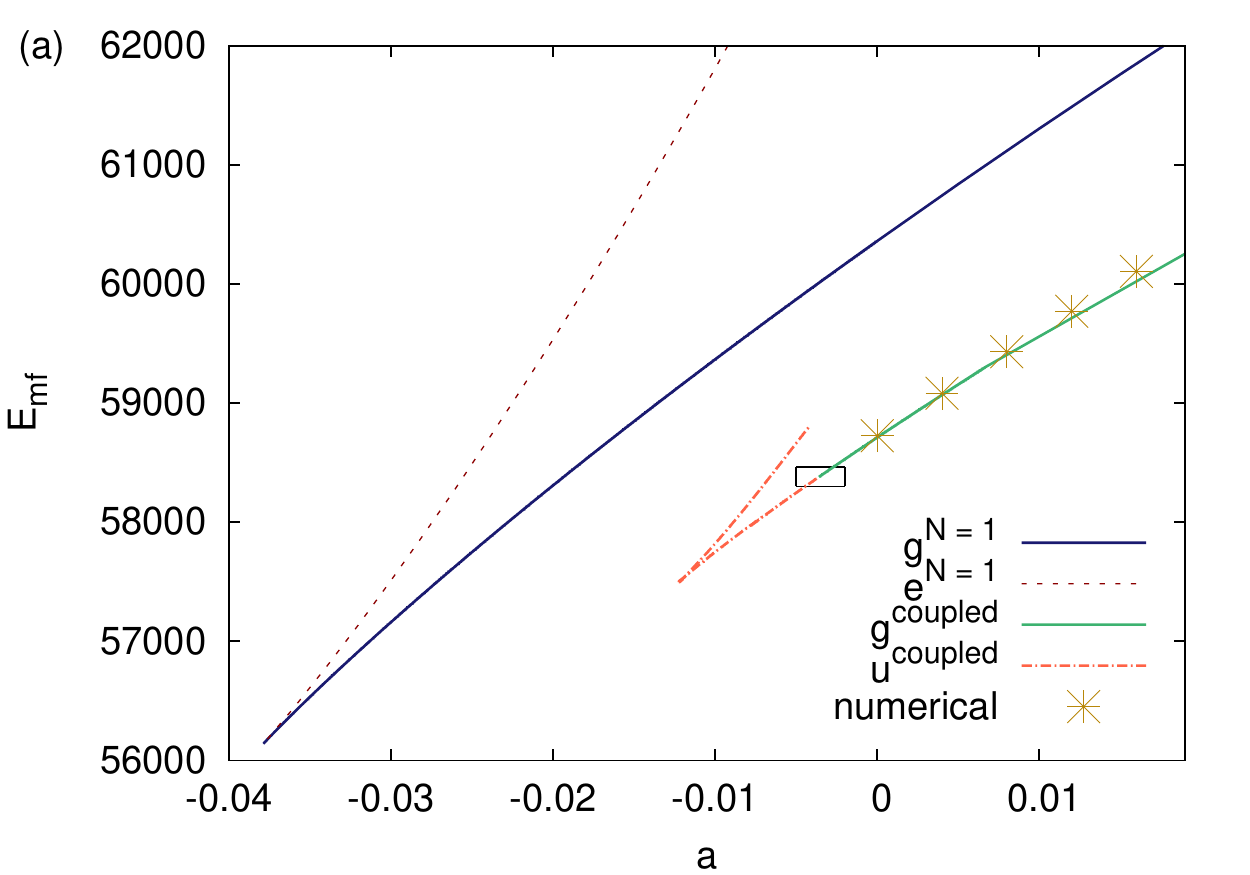}\\
\includegraphics[angle=-90, width=0.9\columnwidth]{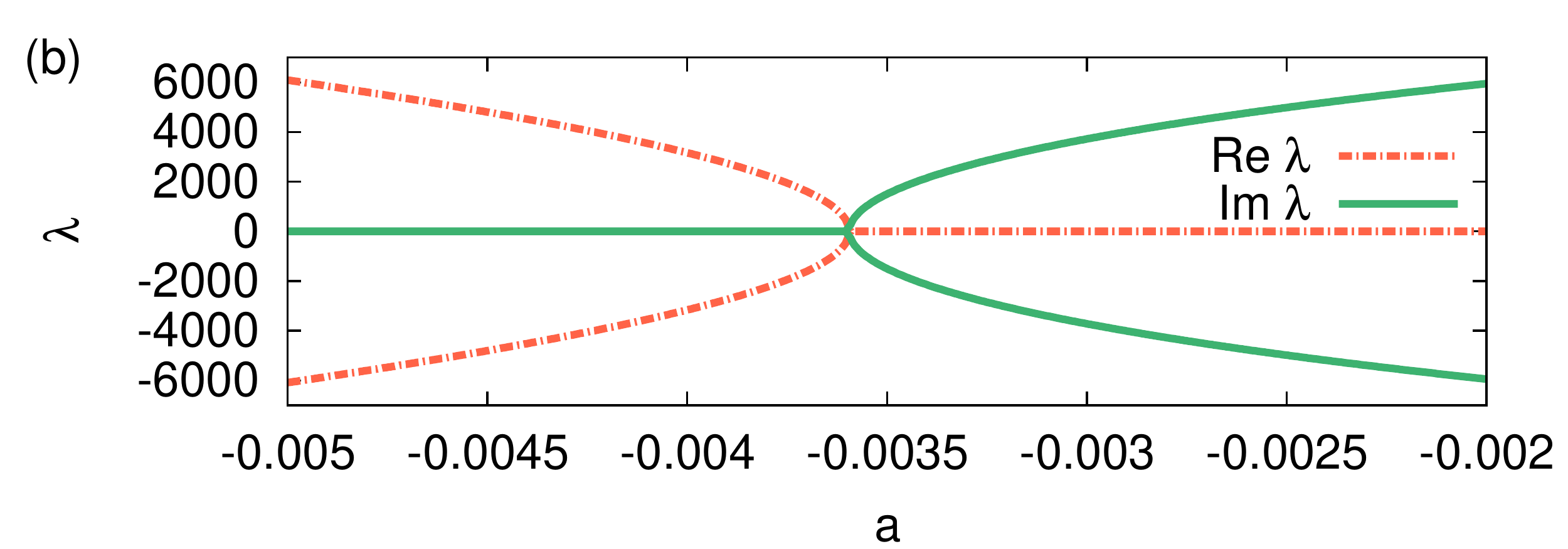}
}
\caption{\label{Emf}  (a) Mean field energy of a dipolar condensate for (particle number scaled) trap frequencies  $N^2\gamma_z = 25200$ and $N^2\gamma_{\rho} = 3600$ as a function of the scattering length. In the variational calculation with
one Gaussian a stable ground state ($\rm g^{N = 1}$) and an unstable excited state ($\rm e^{N = 1}$) emerge in a tangent 
bifurcation. Using coupled Gaussians two unstable states emerge (labeled $\rm u^{coupled}$), of which the lower one 
turns into a stable ground state ($\rm g^{coupled}$) in a pitchfork bifurcation. 
(b) Stability eigenvalues $\lambda$ of the pitchfork bifurcation point for calculations with 6 coupled Gaussians, scattering length in rectangle marked in (a). Real and imaginary parts of two selected eigenvalues of the Jacobian (\ref{Jacobian}) as a function of the scattering length. For $a <  a^{\mathrm{p}}_{\mathrm{cr}} = -0.00359 $ the solution is unstable with one pair of real eigenvalues. At $ a^{\mathrm{p}}_{\mathrm{cr}}$ the real eigenvalues vanish in a pitchfork bifurcation and a stable ground state forms with purely imaginary eigenvalues. Only those eigenvalues involved in the stability change are shown.} 
\end{figure}

Figure~\ref{Emf}~(a) shows, for the same set of trap frequencies, the results for the mean field energy of the condensate as a function of the scattering length 
$a$ (in units $a_d$) for a wave function with one Gaussian, and for 5 coupled Gaussian wave packets. 
Results obtained using 6 Gaussians would be indistinguishable in the figure from those obtained using  5  Gaussians, and the results for 2--4 Gaussians are not shown for the sake of clarity of the figure.

Similar to the above findings for monopolar condensates, and as is known from
previous variational calculations  \cite{gelbPatrick} for dipolar condensates, for $N=1$ two branches of solution are born in a tangent bifurcation. The energetically higher branch
has purely real stability eigenvalues  $\pm \lambda^R$, corresponding to
an unstable excited state $\rm e^{N = 1}$, the lower branch possesses purely imaginary
eigenvalues $\pm \lambda^I$ and corresponds to the stable ground state $\rm g^{N = 1}$. At the bifurcation point the branches of the stability eigenvalues merge and vanish.

The situation is different if the condensate wave function is described by more than one Gaussian. As the scattering length is decreased from positive values towards the tangent bifurcation, the branch corresponding to the ground state $\rm g^{coupled}$ turns into an unstable state $\rm u^{coupled}$ at a
scattering length of $ a^{\mathrm{p}}_{\mathrm{cr}} = -0.00359 $.
This is evident from the stability analysis shown in Fig.~\ref{Emf}~(b)
where the stability eigenvalues for the ground state, calculated using
6 Gaussians, are plotted in a small interval of the scattering length
around  $ a^{\mathrm{p}}_{\mathrm{cr}} $.
Above $ a^{\mathrm{p}}_{\mathrm{cr}}$ the eigenvalues are purely imaginary, 
below they are purely real.
[Note that in a Bogoliubov analysis this instability should appear as a
dynamical instability.]
The ground state remains unstable down to the tangent bifurcation point 
at $ a^{\mathrm{t}}_{\mathrm{cr}} = -0.01224$, where it joins the branch of 
the unstable excited state.

The quality of the calculation using 5 Gaussian wave packets is also demonstrated in Fig.~\ref{Emf}~(a) where the results of a numerically grid calculation by imaginary time evolution are shown by crosses. Evidently
the numerical results and the results obtained using 5 coupled Gaussians 
excellently agree. The imaginary time calculation, however,
can only trace the stable branch of the solution and fails for the unstable branch. Thus it is  demonstrated that the
Gaussian wave packet method is not only numerically accurate but also capable of giving access to regions of the
space of solutions of the Gross-Pitaevskii equation with dipolar interaction that are difficult to investigate by
conventional numerical full quantum calculations. 
 
\begin{figure}[ht!]
\includegraphics[angle=0, width=0.9\columnwidth]{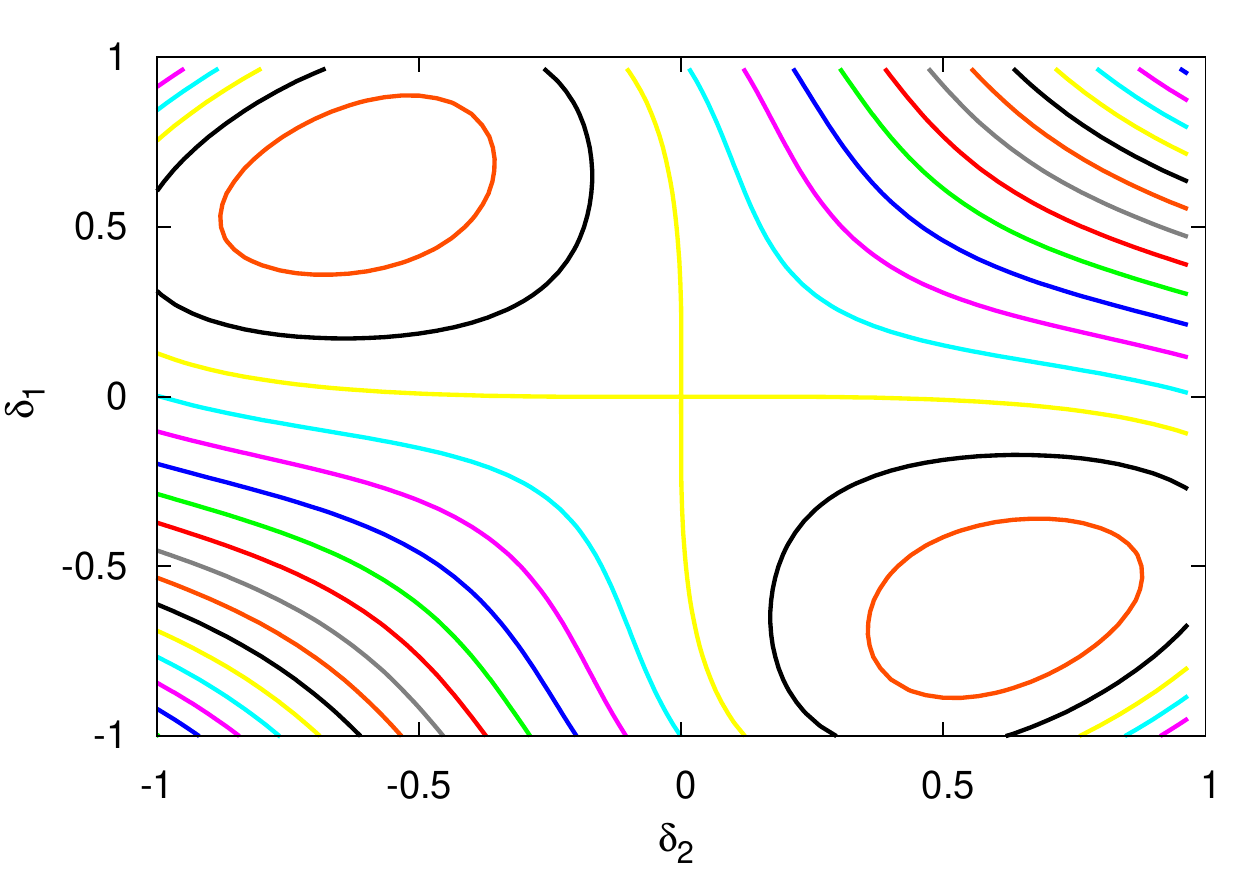}
\caption{\label{phaseportrait} Contour plot of the mean field energy with the eigenvectors corresponding to the eigenvalues of Fig.~\ref{Emf}~(b) linearizing the vicinity of the fixed point ($\delta_1, \delta_2 $ in arbitrary units). The figure shows $a = -0.0036$ close below the pitchfork bifurcation point, showing three fixed points: Two stable and one hyperbolic.}
\end{figure}

The phenomenon  of one smooth  branch of solutions becoming unstable as a function of a control parameter is reminiscent of a {\em pitchfork} bifurcation.  The two stable solutions on the  fork arms which  should also be born in a pitchfork 
bifurcation, and  
exist in a tiny neighborhood $( a^{\mathrm{p}}_{\mathrm{cr}} - \epsilon) < a <  a^{\mathrm{p}}_{\mathrm{cr}}$, are numerically hard to trace and therefore not plotted  in the figure. Their existence, and
the pitchfork type of the bifurcation, however, can be proven by looking at
the ``phase portrait'' plotted in Fig.~\ref{phaseportrait} at a value of the scattering length $a = -0.036$ slightly below $ a^{\mathrm{p}}_{\mathrm{cr}}$.  
Figure~\ref{phaseportrait} shows contours of equal deviation of the mean field
energy from that of the ground state in the plane spanned by
the two eigenvectors whose eigenvalues are involved in the stability change in Fig~\ref{Emf}~(b). The coordinate axes $\delta_1, \delta_2$ correspond to small variations   
of the Gaussian parameters in the eigenvector directions around the hyperbolic 
fixed point solution located  at the origin. The portrait clearly reveals the existence of two nearby elliptic fixed points 
corresponding to two additional stable solutions. Therefore, in a small interval $\epsilon$ of the scattering length 
below $ a^{\mathrm{p}}_{\mathrm{cr}}$, $ ( a^{\mathrm{p}}_{\mathrm{cr}} - \epsilon) < a <  a^{\mathrm{p}}_{\mathrm{cr}}$, there exist two additional branches, besides the unstable solution. This proves that the bifurcation is of pitchfork type.
Note that the classification of the condensate as unstable for $a <  a^{\mathrm{p}}_{\mathrm{cr}}$ nevertheless remains true in physical terms due to the numerically small value 
of $\epsilon$. We also note that for $a >  a^{\mathrm{p}}_{\mathrm{cr}}$ the phase portrait possesses 
only one 
elliptic fixed point cooresponding to the stable stationary ground state.

Is there a chance of observing dipolar BECs on the stable fork arms?
The answer probably is no, in the same way as it is in the case of the question
of observing the transition to structured ground states, possibly associated
with a roton instability, shortly before collapse. The reason is
the difficulty of adjusting trap frequencies and the scattering length
to the necessary precision in a real experiment. Nevertheless theoretical
investigations of this type close to the threshold of instability of dipolar 
condensates are valuable in their own right since they help to understand 
the nature of the collapse, and thus of ``what's really going on''.


\end{document}